\documentclass[]{article}
\usepackage[a4paper, total={6in, 8in}]{geometry}
\usepackage[utf8]{inputenc}
\usepackage{amsmath}
\usepackage{bibref}
\usepackage{graphicx}
\usepackage{makecell}
\usepackage{tabularray}
\usepackage{multicol}
\usepackage{subcaption}
\DeclareMathOperator{\sech}{sech}
\title{Observational evidence for parametrized emergent dark energy models}
\author{Sarath Nelleri\footnote{sarathn@iitk.ac.in} and Navaneeth Poonthottathil\footnote{navaneeth@iitk.ac.in} \\ \small Department of Physics, Indian Institute of Technology Kanpur, Kalyanpur 208016, India}

\begin{document}
\date{}
\maketitle

\begin{abstract}
	Recent cosmological observations show a statistically significant tension in the estimated values of the cosmological parameters within the standard $\Lambda$CDM framework. In a recent study, Li and Shaﬁeloo introduced a simple Phenomenological Emergent Dark Energy (PEDE) model, which possesses the same number of parameters as that of the $\Lambda$CDM model. Their research highlighted this model as a viable alternative to $\Lambda$CDM, capable of alleviating the Hubble tension and explaining the late-time cosmic acceleration. Following this, we consider a series of PEDE-type models where a new parameter $b$ is introduced in the dark energy expression that distinguishes one model from the other and is designated as bPEDE models. The PEDE and $\Lambda$CDM models were the special cases of the bPEDE model. In contrast to the PEDE model, the bPEDE model demonstrates the presence of dark energy in the past while indicating its absence in the asymptotic future. Confronting these models with the observational Hubble data (OHD) shows that a series of bPEDE models fit the data better than the PEDE model and the standard $\Lambda$CDM model. Notably, the Hubble constant ($H_0$) value computed using the best-fit bPEDE models closely aligns with the CMBR prediction. It significantly deviates from the local measurement at a significance level of approximately $ 3.4\sigma$ for the model independent OHD data combination. The outcome suggests reconsidering the systematic uncertainties associated with the local measurement. The best-fit bPEDE models predict the deceleration to acceleration transition at a redshift $z_T \sim 0.78$ which is in close agreement with the $\Lambda$CDM prediction. The age of the universe predicted by the bPEDE model is $\sim 14$ Gyr, slightly higher than the age predicted by the $\Lambda$CDM model. The statefinder trajectory reveals a quintessence nature of dark energy. Our analysis shows that the best-fit bPEDE model outperforms the  PEDE and $\Lambda$CDM models and hint towards a static future of the universe. 

\end{abstract}

\section*{Introduction}\label{intro}
Observations on Type Ia supernovae indicated that the present evolution of the universe is accelerating\cite{perlmutter1999measurements, riess1998observational, perlmutter1999constraining}. It is well-supported by comprehensive observational probes such as Cosmic Microwave Background Radiation (CMBR)\cite{mather1990preliminary, mather1994measurement, tegmark2003high, hinshaw2013nine, ade2016planck, aghanim2020planck}, the Baryon Acoustic Oscillation (BAO)\cite{beutler20116df, aubourg2015cosmological}, the Large Scale Structure (LSS)\cite{efstathiou1992cobe, tegmark2004cosmological, alam2017clustering, beutler20116df} and Observational Hubble Data (OHD)\cite{ma2011power, hernandez2020generalized, sudharani2023hubble}. The accelerated expansion of the universe during the late phase can be modelled by combining the cosmological constant that produces negative pressure with the cold dark matter, known as the $\Lambda$CDM model\cite{baumann2022cosmology, coles2003cosmology, padmanabhan2010gravitation, ryden2017introduction}. As a result of its remarkable success in explaining the observational data, the $\Lambda$CDM model is considered to be the standard model of cosmology. Investigation into Cosmic Microwave Background Radiation (CMBR) and local measurement of Hubble parameter with respect to redshift highlight a statistically significant tension regarding the present value of the cosmological parameters\cite{di2021realm, schoneberg2019bao+, aloni2022step, alestas2021late, mortsell2018does}. For instance, the estimation of $H_0$ by the Planck collaboration 2018 yielded a value of $67.4\pm 0.5$ km $\text{s}^{-1}$ $\text{Mpc}^{-1}$\cite{aghanim2020planck} whereas the observation on Cepheids by the SHOES collaboration resulted in $74.03\pm 1.42$ km $\text{s}^{-1}$ $\text{Mpc}^{-1}$\cite{riess20162, riess2021cosmic}. This discrepancy showcases the tension in the precise measurement of the Hubble constant, where the tension is about $4.2\sigma.$ In this context, the Planck result is acquired under the assumption of $\Lambda$CDM model. Conversely, the SHOES result is obtained in a model-independent manner. The measurements obtained from the Cosmic Microwave Background (CMB) exhibit high precision and align effectively with the predictions of the standard cosmological model. However, the researchers mostly posit that the discrepancy lies within the $\Lambda$CDM model, refuting the possibility of systemic errors in the local measurements. Furthermore, the parameter that quantifies the matter fluctuation amplitude ($S_8$) obtained from CMBR data is $S_8 = 0.832\pm 0.013$\cite{aghanim2020planck} and those obtained from galaxy clustering and weak lensing is $S_8 = 0.762^{0.025}_{-0.024}$\cite{abbott2018dark}, the discrepancy is at a level of $\sim 2\sigma$. Many attempts have been made to resolve these issues. These discrepancies may shed light on the most compelling evidence of physics beyond the standard cosmology. Various attempts have been made to resolve these difficulties. For instance, ref. \cite{poulin2019early} presents early dark energy that behaves like a cosmological constant in the early universe, alleviating the Hubble tension. Other notable examples comprise phantom dynamical dark energy models\cite{dahmani2023smoothing}, negative dark energy models\cite{sen2023cosmological, malekjani2023negative}, the dissipative axion particle model\cite{berghaus2020thermal}, baryon inhomogeneities resulting from primordial magnetic fields\cite{jedamzik2020relieving}, modified gravity models\cite{adi2021can} and many more. Nevertheless, certain models exhibit dynamic instability, while support for other models in contrast to $\Lambda$CDM, considering observational data, is limited. Recent studies show that the cosmological tension problem entails a redshift evolution of the cosmological parameters\cite{malekjani2023negative}. 

%In ref.\cite{li2019simple}, Li and Shafieloo proposed a simple (zero degrees of freedom) phenomenological emergent dark energy model (PEDE) that alleviates the Hubble tension. 
Phenomenological emergent dark energy (PEDE) model\cite{li2019simple} has been proposed as a potential alternate to the $\Lambda$CDM model with a motivation of alleviating the Hubble tension. As described within the framework of this model, dark energy has no effective presence in the past and only emerges as time advances. In this model, the dark energy density is assumed to have the form
\begin{align}
	\label{eq:PEDE}
	\Omega_{D} = \Omega_{D_0}[1-\tanh(\log_{10}(1+z))],
\end{align}
where $\Omega_{D_0}$ is the dark energy density at present and $z$ is the redshift. In Eq. (\ref{eq:PEDE}), authors preferred the base of the logarithm to ten. Even if one assume that the evolution of the dark energy follows a hyperbolic tangent of a logarithmic function, there is no theoretical reason to set the base of the logarithm to ten. Such an assumption leaves the possibility of a model to have a different base that provides a better fit to the data than the PEDE model and the standard $\Lambda$CDM model. The PEDE model, as described in ref. \cite{li2019simple} highlighted that the analysis relying on the data sets encompassing SNIa, BAO, Ly$\alpha$ BAO and CMBR prefer the PEDE model substantially better than the $\Lambda$CDM model. The authors employed a hard-cut $2\sigma$ lower bound prior for the $H_0$ in the analysis to get this better evidence, a choice which remarkably influenced their findings. Under this assumption, their analysis demonstrated a substantial reduction of the Hubble tension, assuming the reliability of the local measurement and effectively refuting the possibility of systematic errors in the local measurement. It is conclusive from their analysis that, with the absence of any $H_0$ prior, the $\Lambda$CDM model gains better preference over the PEDE model. Notably, the $H_0$ value obtained from the $\Lambda$CDM model closely aligns with the value obtained by the Planck measurement when considering the pantheon+BAO data combinations. Despite these challenges, the PEDE-type evolution of dark energy presents a promising candidate to explain the late-phase acceleration of the universe and gain deeper insights into the tension problem within cosmology.
%Furthermore, the analysis presented in ref.\cite{li2019simple} shows that the $\Lambda$CDM model is preferred over the PEDE model in the absence of prior information of the Hubble constant. An interesting part of their analysis is that the PEDE model is outperforming the $\Lambda$CDM model for various cosmological observations such as SNIa, BAO, $\text{Ly}\alpha$ BAO, and CMBR. There is, however, a hard-cut $2\sigma$ lower bound prior for $H_0$ from the recent local measurement is assumed, which is not an appropriate choice of prior since the tension between the CMBR measurement and the local measurement is more than $4\sigma$. In their analysis, the local $H_0$ measurement by the Supernovae H0 for the Equation of State (SHOES) collaboration \cite{riess2019large, riess20162, riess2018type} is assumed to be accurate and completely avoiding the estimated value of $H_0$ by the Planck collaboration.
%There are two fundamental reasons why the CMBR data is considered the most robust probe of the early universe: first, the error bar in the measurement is very small, and second, the $\Lambda$CDM model fits the data very well. Therefore, it is not reasonable to perform a data analysis in which the CMBR prediction is not considered while considering the prior. 

Attempts has been made to generalize the PEDE model including extra parameters, the resulting models are known as Generalized Emergent Dark Energy (GEDE) models\cite{hernandez2020generalized, li2020evidence, yang2021generalized, rezaei2020bayesian, benaoum2022modified}. Indeed, these models posit a more generalized form of dark energy, $\Omega_{D}(z) = \Omega_{D_0} \times \frac{F(z)}{F(z=0)}$ with $F(z) = 1 - \tanh([\log_{10}(1+z) - \log_{10}(1+z_t)])$, where $z_t$ is the transition redshift. Investigation of the GEDE models utilizing the observational probes such as CMBR, BAO, Type Ia supernovae and $H_0$ from Hubble Space Telescope (HST) indicates a preference for these models over the $\Lambda$CDM model. However, this preference is observed at a relatively modest statistical significance of $2\sigma$\cite{yang2021generalized}. Furthermore, Bayesian inference of PEDE model based on CMBR data demonstrated that the $\Lambda$CDM model provides stronger evidence compared to the PEDE model \cite{rezaei2020bayesian}. Recently, Benaoum et al. \cite{benaoum2022modified} analyzed a modified version of the PEDE model named as Modified Emergent Dark Energy model (MEDE) where a new parameter $\alpha$ is introduced and consequently the dark energy density has the form $\Omega_{D} = \Omega_{D_0}[1-\tanh(\log_{10}(1+z)^{\alpha})]$. The analysis shows that the Hubble tension problem is not completely resolved but somewhat alleviated.

In this study, we focus on PEDE-type dark energy models designated as parmetrized emergent dark energy models (bPEDE) that possess a minimum number of parameters as the $\Lambda$CDM model. We explore the possibility of parametrized emergent dark energy model whose logarithm base can be adjusted to arbitrary numbers, that fits the observational data better than PEDE models. Following the comprehensive comparative analysis involving generalized emergent dark energy model, PEDE model and the $\Lambda$CDM model based on the observational Hubble data as presented in ref.\cite{hernandez2020generalized}, we proceed to employ the observational Hubble data for our analysis.  In this study, we are primarily interested in estimating the Hubble constant value assuming the best-fit bPEDE model and assessing whether the model prefers the Planck collaboration's measurement of the CMBR or SHOES's local measurement. In addition, we test the performance of the bPEDE model against the standard $\Lambda$CDM model and the PEDE model adopting the Bayesian statistics to determine which model is preferred by the OHD data. We further analyze the background evolution of the universe within the framework of the bPEDE models.

This paper is organized as follows. In section 2, we discuss the bPEDE-type models. In section 3, we perform the parameter inference and model selection using the observational Hubble data. In section 4, we study the evolution of cosmographic parameters. We conclude in section 5.

\section*{Parametrized emergent dark energy (bPEDE) models}\label{model} 
Observations suggest that the phenomenological emergent dark energy model (PEDE) is a potential alternative explanation for the accelerated expansion of the universe and resolves the Hubble tension problem. According to the PEDE model, dark energy is a dynamical quantity, and its dynamical behaviour is described in Eq. (\ref{eq:PEDE}). Our approach extends the PEDE model to allow any positive real number to be the base of the logarithmic function. In this context, we may refer to these models as bPEDE models, where `b' represents the base of the logarithmic function. The functional form of the dark energy density is considered the same. At the same time, base `b' differentiates one model from the other so that all the models under consideration are very close to the PEDE model. The evolution of the dark energy density in the bPEDE model has the form given by,
\begin{align}
	\label{eq:bPEDE}
	\Omega_{D} = \Omega_{D_0}[1-\tanh(\log_{b}(1+z))],
\end{align}
where $\Omega_{D_0}$ is the present value of the dark energy density, and $z$ is the redshift related to the scale factor ($a$) as $1+z = 1/a$. The matter density and dark energy density satisfy individual conservation equations,
\begin{align}
	\label{eq:C1}
	\dot{\rho_m} + 3H\rho_m = 0, \\
	\label{eq:C2}
	\dot{\rho_D} + 3H(1+\omega_D)\rho_D =0,
\end{align}
where $\omega_D$ is the equation of state of dark energy density. The progress of $\omega_D$ with redshift is obtained from Eq. (\ref{eq:C2}), can be expressed as 
\begin{align}
	\label{eq:omega1}
	\omega_D(z) = \frac{1}{3}\frac{d\ln \Omega_D}{dz}(1+z) - 1,
\end{align}
where $\Omega_D$ is the dark energy density normalized over the present value of the critical density, $\rho_{c_0} = 3H_0^2/8\pi G$. Substituting Eq. (\ref{eq:bPEDE}) in (\ref{eq:omega1}), the evolution of $\omega_D$ is obtained as 
\begin{align}
	\label{eq:omega2}
	\omega_D(z) = -\frac{1}{3\ln b}\left(1 + \tanh[\log_{b}(1+z)]\right)-1.
\end{align}
The eq. (\ref{eq:omega2}) shows that the nature of dark energy depends on the base `b'. For $0<b<\frac{1}{e}$, the model is a quintessence-type dark energy model, while for $1<b<\infty$, the model resembles a phantom dark energy model. At present, $\omega_D (z=0) = \frac{-1}{3\ln b}-1$, shows that the present value of the equation of state of dark energy depends on the value of `b'. The evolution of the matter density is obtained by solving the Eq. (\ref{eq:C1}), expressed as $\rho_m = \rho_{m_0}(1+z)^3$, where $\rho_{m_0}$ is the matter density at present. The evolution of matter density is same as that of the PEDE and $\Lambda$CDM models. The evolution of the Hubble parameter within a flat Friedmann-Lemaitre-Robertson-Walker (FLRW) metric\cite{peebles1993principles} is
\begin{align}
	\label{eq:H}
	H^2(z) = H_0^2\left[\Omega_{m_0}(1+z)^3 + \Omega_{D_0}[1-\tanh(\log_{b}(1+z))] \right]
\end{align}
At present, $z=0$, the Hubble parameter ($H$) reduces to $H_0$, where $H_0$ is the present value of the Hubble parameter. In the far past, $z\rightarrow \infty$, the evolution of dark energy density depends on the value of parameter `b'. In the asymptotic future, $z\rightarrow -1$ the matter density, $\Omega_m \rightarrow 0$. In contrast the value of $\Omega_D$ depends on the value of `b', consequently influencing the asymptotic evolution of the Hubble parameter.

\section*{Parameter inference and model Selection}\label{data}
The parameter estimation for the Phenomenological Emergent Dark Energy (PEDE) model, as presented in the reference \cite{li2019simple}, indicates that the observational data, including the Type Ia supernovae, the Cosmic Microwave Background Radiation (CMBR), and the Baryon Acoustic Oscillations (BAO), without the assumption of a prior for the Hubble constant ($H_0$), tend to favour a value of $H_0$ closely aligned with the measurement obtained by the SHOES collaboration, approximately $\sim 74$. However, it is important to note, based on their analysis without assuming any hard-cut prior on $H_0$, that both the $\Lambda$CDM model and the Chevallier-Polarski-Linder (CPL) parameterization model \cite{chevallier2001accelerating, linder2003exploring} show better Deviance Information Criterion (DIC) \cite{spiegelhalter2002bayesian, liddle2007information} compared to the PEDE model. Intriguingly, these models yield a value of $H_0$ in close agreement with the predicted CMBR value, around $\sim 67$, for data combinations involving the Pantheon supernova compilation along with BAO, and either Lyman-alpha and CMB data.
Notably, the $\Lambda$CDM and CPL parameterization models predict a value of $H_0$ close to the CMBR prediction, regardless of whether the data set includes CMBR data or not. The authors observed that the value of $H_0$ aligns closely with the local measurement value when assuming $1\sigma$ or $2\sigma$ priors for $H_0$ taken from the SHOES result. Under these conditions, the PEDE model shows better evidence than the $\Lambda$CDM model\cite{li2019simple}. In this context, this study aims to explore the bPEDE models presented in eq. (\ref{eq:H}), aiming to achieve a better fit than the PEDE and $\Lambda$CDM models and to examine the predicted value of $H_0$ within the bPEDE model.

In this study, as presented in ref. \cite{yang2021generalized}, we adopt observational Hubble data (OHD) to perform both parameter inference and model selection. The OHD dataset comprises 51 Hubble parameter values observed within the redshift range of $0.07<z<2.36$. Among these, 31 data points are obtained model-independently from the differential age (DA) technique \cite{jimenez2002constraining, moresco2012improved}. The remaining 20 data points are obtained through Baryon Acoustic Oscillations (BAO) measurements, where assumptions based on standard cosmology ($\Lambda$CDM) are used to estimate the sound horizon at the drag epoch \cite{magana2018cardassian}. Hence, these data points may introduce some biased constraints for the model parameters.

Nevertheless, to avoid potential biases, homogenized and model-independent OHD data is presented in reference \cite{magana2018cardassian}, obtained by employing the sound horizon at the drag epoch from Planck collaboration 2016 data. Within the scope of this study, we consider three distinct data combinations: Data1: OHD (DA) + OHD (homogeneous from BAO),  Data2: OHD (DA) + OHD (non-homogeneous from BAO) and Data3: OHD (DA) for our analyses. We adopt Bayesian statistics for the parameter inference and model selection. Bayesian statistic is based on the Bayes theorem, which provides a gratifying description to obtain the posterior distribution of the model parameters ($\theta$) for a given set of data (D) and model (M)\cite{john2002comparison, trotta2008bayes}. According to Bayes theorem, 
\begin{align}
	\label{eq:B}
	P(\theta|D,M) = \frac{P(D|\theta,M)P(\theta|M)}{P(D|M)},
\end{align}
where $P(\theta|D,H)$ is the posterior distribution of the model parameter, $P(D|\theta,M)$ is the likelihood, $P(\theta|H)$ is the prior and $P(D|M)$ is just a normalization factor that represents the evidence of the model. It evidence is irrelevant for the parameter estimation. However, it is the central quantity of interest when we do the model selection. The prior probability encapsulates any information available regarding the model parameters before acquiring the data\cite{hobson2010bayesian, verde2010statistical}. The choice of prior is contingent upon any information we have regarding the model and depends on the quality of judgment\cite{padilla2021cosmological}. Nonetheless, once the prior is established, successive application of Bayes' theorem results in convergence towards a common posterior. The likelihood is defined as
\begin{align}
	\label{eq:L}
	P(D|\theta,M) \equiv \exp(-\chi^2(\theta)/2),
\end{align}
where the $\chi^2$ is defined as
\begin{align}
	\label{eq:chi}
	\chi^2(\theta) = \sum_{k}^{}\left[\frac{H_k - H_k(\theta)}{\sigma_k}\right]^2
\end{align}
Here, $H_k$ is the Hubble parameter value corresponding to the redshift value $z_k$ given in the OHD data, $H_k(\theta)$ is the corresponding theoretical value obtained from the model and $\sigma_k$ is the standard deviation in the measured values. Marginalizing over all other parameters except the parameter of interest, we obtain the posterior distribution of the parameter of interest\cite{trotta2008bayes, rezaei2021comparison, rezaei2020bayesian}. For instance, if the model has the parameter space $\theta = (\theta_1, \theta_2.....\theta_n)$, the marginal probability of $\theta_1$ can be expressed as
\begin{align}
	\label{eq:marginal}
	p(\theta_1|D,M) = \int p(\theta|D,M)d\theta_2...d\theta_n,
\end{align}
which represents a one-dimensional posterior distribution of the model parameter $\theta_1.$ A two-dimensional posterior can also be defined similarly. We use the Markov Chain Monte Carlo (MCMC) method for the numerical simulation.
\begin{table}
	\centering
	\caption{Jeffrey's scale}
	\begin{tabular}{ |c | c |c | }
		\hline
		$|\ln B_{01}|$ & Probability & Strength of evidence \\  \hline
		$< 1.0$ & $<0.750$ & Inconclusive \\ \hline
		$1.0$ & $0.750$ & weak evidence \\ \hline
		$2.5$ & $0.923$ & Moderate evidence \\ \hline
		$5.0$ & $0.993$ & Strong evidence    \\ \hline
	\end{tabular}
	\label{tab:1}
\end{table}

Bayesian evidence $p(D|M)$ plays a central role in the model selection. It is obtained by taking the average of likelihood over the prior for a particular model of choice, which can be expressed as
\begin{align}
	\label{eq:evidence}
	p(D|M) = \int d\theta p(D|\theta, M)p(\theta|M)
\end{align}
The evidence of one model ($M_0$) over the other ($M_1$) is quantified using the Bayes factor ($B_{01}$), which is defined as the ratio of Bayesian evidence of the models, expressed as
\begin{align}
	\label{eq:bayes}
	B_{01} \equiv \frac{p(D|M_0)}{p(D|M_1)}.
\end{align}
The empirical scale for quantifying the strength of evidence is called Jeffrey's scale, presented in Tab. \ref{tab:1}\cite{trotta2008bayes}. 
We also use information criteria, which are frequently used in cosmology for the model selection, such as Akaike Information Criterion (AIC)\cite{sakamoto1986akaike, tan2012reliability, rezaei2021comparison} and Bayesian Information Criterion (BIC)\cite{liddle2004many, arevalo2017aic}. The Akaike Information Criterion (AIC), which is essentially a frequentist criterion that includes a penalty term equal to twice the number of parameters present in the model ($k$), is defined as
\begin{align}
	\label{AIC}
	AIC \equiv -2\ln \mathcal{L}_{max} + 2k,
\end{align}
where $\mathcal{L}_{max} \equiv p(D|\theta_{max}, M)$ is the maximum likelihood value. The Bayesian Information Criterion (BIC), which is also known as the Schwarz Information Criterion, that follows from a Gaussian approximation to the Bayesian evidence in the limit of larger sample size is defined as
\begin{align}
	\label{AIC}
	BIC \equiv -2\ln \mathcal{L}_{max} + k\ln N,
\end{align}
where $N$ is the number of data points. The model that minimizes AIC and BIC is considered as the best model.

\begin{figure}
	\centering
	\begin{subfigure}[b]{0.32\textwidth}
		\centering
		\includegraphics[width=\textwidth]{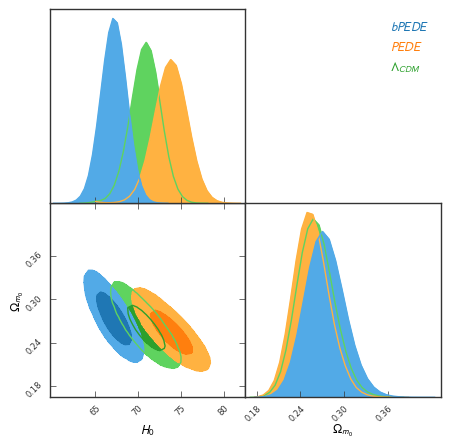}
		\caption{OHD (DA+Homogenous)}
		\label{fig:homoA}
	\end{subfigure}
	\hfill
	\begin{subfigure}[b]{0.32\textwidth}
		\centering
		\includegraphics[width=\textwidth]{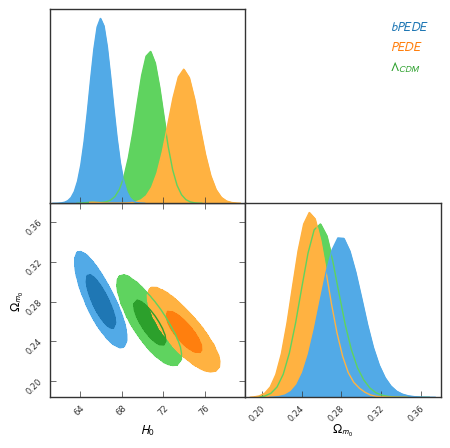}
		\caption{OHD (DA+Non-homogenous)}
		\label{fig:NHA}
	\end{subfigure}
	\hfill
	\begin{subfigure}[b]{0.32\textwidth}
		\centering
		\includegraphics[width=\textwidth]{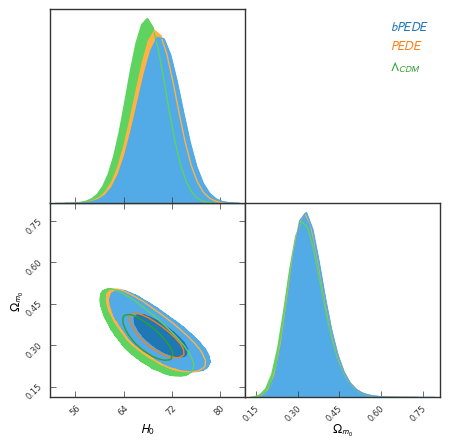}
		\caption{OHD (DA)}
		\label{fig:DAA}
	\end{subfigure}
	\caption{The 1-D and 2-D marginal likelihood of the model parameters of the bPEDE model for the different OHD data combinations.}
	\label{fig:AIC1}
\end{figure}

\begin{figure}
	\centering
	\begin{subfigure}[b]{0.325\textwidth}
		\centering
		\includegraphics[width=\textwidth]{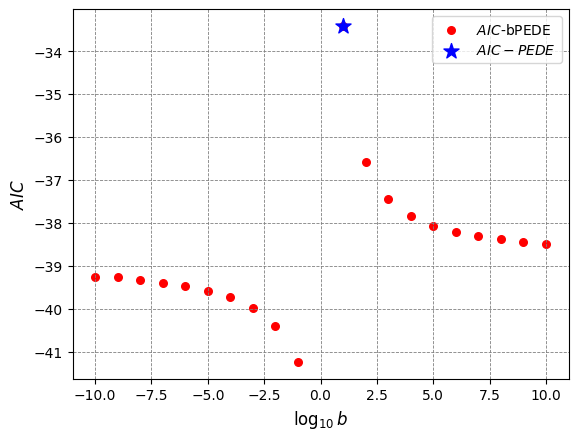}
		\caption{OHD (DA+Homogeneous)}
		\label{fig:homo}
	\end{subfigure}
	\hfill
	\begin{subfigure}[b]{0.325\textwidth}
		\centering
		\includegraphics[width=\textwidth]{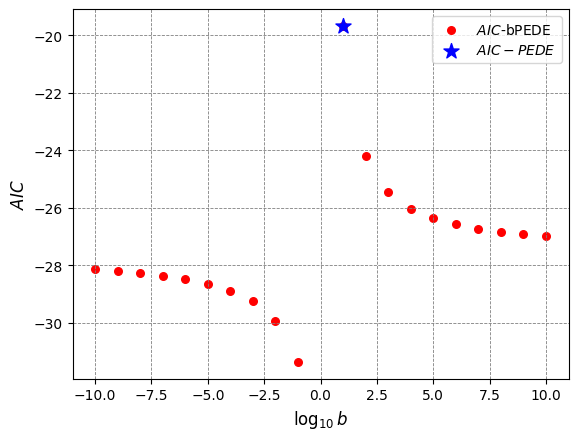}
		\caption{OHD (DA+Non-homogeneous)}
		\label{fig:NH}
	\end{subfigure}
	\hfill
	\begin{subfigure}[b]{0.33\textwidth}
		\centering
		\includegraphics[width=\textwidth]{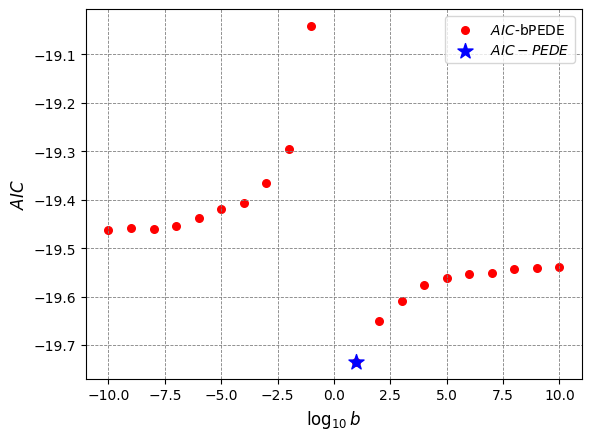}
		\caption{OHD (DA)}
		\label{fig:DA}
	\end{subfigure}
	\caption{Variation of AIC against $\log_{10}b$ for bPEDE models for the different OHD data combinations. The star (blue) represents the AIC corresponding to the PEDE model}
	\label{fig:AIC}
\end{figure}

\begin{figure}
	\centering
	\begin{subfigure}[b]{0.32\textwidth}
		\centering
		\includegraphics[width=\textwidth]{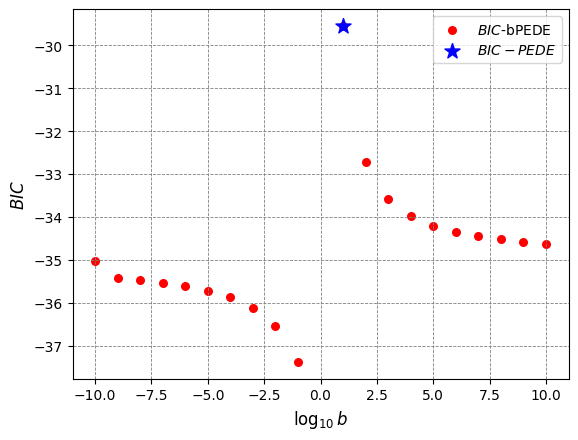}
		\caption{OHD (DA+Homogeneous)}
		\label{fig:homo1}
	\end{subfigure}
	\hfill
	\begin{subfigure}[b]{0.32\textwidth}
		\centering
		\includegraphics[width=\textwidth]{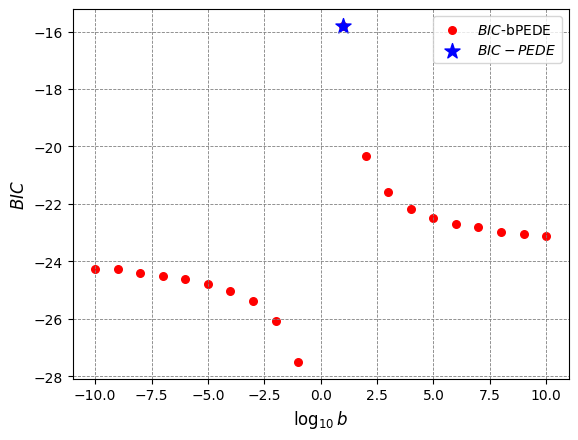}
		\caption{OHD (DA+non-homogeneous)}
		\label{fig:NH1}
	\end{subfigure}
	\hfill
	\begin{subfigure}[b]{0.325\textwidth}
		\centering
		\includegraphics[width=\textwidth]{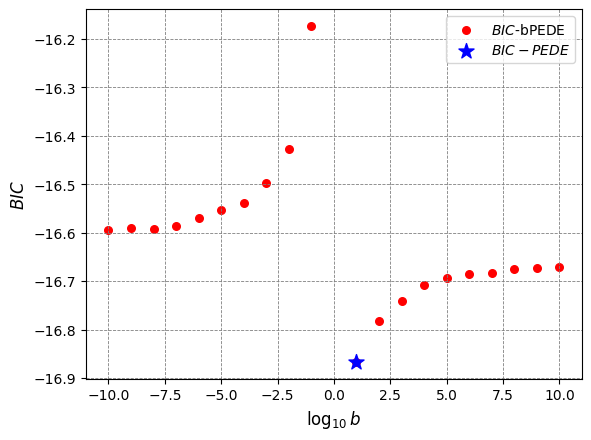}
		\caption{OHD (DA)}
		\label{fig:DA1}
	\end{subfigure}
	\caption{Variation of BIC against $\log_{10}b$ for bPEDE model for the different OHD data combinations. The star (blue) represents the BIC corresponding to the PEDE model}
	\label{fig:BIC}
\end{figure}

Initially, we are interested to see how the AIC and BIC change according to the change in base ($b$) of the logarithmic function presented in Eq. (\ref{eq:bPEDE}). We consider the range of $\log_{10} b$ between $-10$ and $+10$. The variation of AIC and BIC with respect to $\log_{10}b$ is presented in Fig. (\ref{fig:AIC}) and (\ref{fig:BIC}), respectively. It is evident from the figures that there is a series of bPEDE models that gives a better fit to all the data combinations compared to the PEDE model ($\log_{10}b = 1$). The model that gives minimum AIC and BIC is expected to be in the range $-1<\log_{10}b<0$ for Data1 and Data2 while $0<\log_{10}b<1$ for Data3. We computed the AIC and BIC for the PEDE model with base $-1<\log_{10}b<0$ for the Data1 and Data2 and $0<\log_{10}b<1$ for the Data3 and found that the best-fit model corresponds to $b=-0.7, -0.6$ and $0.7$ for the Data1, Data2 and Data3 respectively. A comparison between the $\chi^2_{min}$, AIC, BIC and the model parameters of the bPEDE model, PEDE model and the standard $\Lambda$CDM model is presented in Tab. \ref{tab:chi2}. Also, the 1-D and 2-D posterior distributions of the model parameters are presented in Fig. \ref{fig:AIC1}. 
The Tab. \ref{tab:chi2} shows that the bPEDE model gives lower AIC and BIC values than the PEDE and the $\Lambda$CDM models. It indicates that observational Hubble data prefer the bPEDE model over the PEDE and $\Lambda$CDM models. More interestingly, the bPEDE model that best fits these data combinations predicts the value of $H_0$ close to the value obtained from CMBR data. None of these models prefers the value of $H_0$ close to the one obtained by the SHOES collaboration. The PEDE model gives values of $H_0$ close to the value obtained by SHOES collaboration for all the data combinations. However, it should be noted that it is possible to construct a model having PEDE-like behaviour that gives a better fit than the PEDE model, which gives the value of $H_0$ close to the CMBR predicted value. The $\Lambda$CDM model gives better-fit to Data1 and Data2 as compared to the PEDE model, and it gives the value of $H_0$ in between the CMBR predicted value and the one obtained by the SHOES collaboration. The best-fit bPEDE, PEDE and $\Lambda$CDM give almost similar fit to the OHD (DA) data, and suggest the CMBR predicted value of $H_0$ for the OHD (DA) dataset.

Further, we have computed the Bayesian evidence of the best-fit bPEDE, PEDE and the $\Lambda$CDM models using Eq. (\ref{eq:evidence}). The Bayes factor that quantifies the relative evidence is obtained using Eq. (\ref{eq:bayes}). The Bayes factors obtained for the bPEDE model against the $\Lambda$CDM model and the PEDE model are presented in Tab. \ref{tab:3}. It shows that the bPEDE model is preferred over the PEDE model for OHD (DA) + OHD (Homogenous) and OHD (DA) + OHD (Non-homogenous) data combinations with $>75\%$ and $>92.3\%$ probabilities while preferred over the $\Lambda$CDM model $\sim 75\%$ probabilities. All three models give almost the same fit to the OHD (DA) data set. We can conclude that the improved data set prefers the bPEDE model as compared to the PEDE model, and the whole analysis shows moderate evidence of the bPEDE model over the PEDE model, and the evidence against the $\Lambda$CDM is not strong. 

Recently, Almada et al.\cite{hernandez2020generalized} did a comparative study of the PEDE model and the generalized version of the PEDE model called the Generalized Emergent Dark Energy Model (GEDE) in the light of the observational Hubble data set presented in this work. Our analysis based on the AIC and BIC criteria shows that the bPEDE model is preferred over the Generalized Emergent Dark Energy (GEDE)  model presented in ref.\cite{hernandez2020generalized}. The present analysis based on the observational Hubble data shows that there exists a possibility of the bPEDE model having a specific b value that fits better than PEDE, GEDE and the $\Lambda$CDM models. Also, this best-fit model predicts the value to $H_0$ very close to the $H_0$ value obtained from CMBR assuming the standard $\Lambda$CDM model.

\begin{table}
	\centering
	\caption{Comparison between $\chi^2$, AIC, BIC and the model parameters predicted by bPEDE model, PEDE model and the $\Lambda$CDM model.}
	\begin{tabular}{|c|c|c|c|c|c|c|c|}
		\hline 
		Model&Data&$\log_{10}b$&$\chi^2_{min}$&AIC&BIC&$H_0$&$\Omega_{m_0}$\\
		\hline
		&Data1&$-0.7$&$20.90$&$-41.47$&$-37.61$&$67.13\pm 1.42$&$0.2728\pm 0.0025$\\
		\Xcline{2-8}{0.2pt}
		bPEDE&Data2&$-0.6$&$25.16$&$-32.02$&$-28.16$&$65.97\pm 1.01$&$0.2800\pm 0.0196$\\
		\Xcline{2-8}{0.2pt}
		&Data3&$0.7$&$14.40$&$-19.76$&$-16.89$&$69.92\pm 3.41$&$0.3340\pm 0.0572$\\
		\cline{1-8}
		&Data1&$10$&$24.48$&$-33.41$&$-29.55$&$73.81\pm 1.83$&$0.2541\pm 0.0229$\\
		\Xcline{2-8}{0.2pt}
		PEDE&Data2&$10$&$32.06$&$-19.67$&$-15.80$&$73.92\pm 1.37$&$0.2496\pm 0.0170$\\
		\Xcline{2-8}{0.2pt}
		&Data3&10&$14.41$&$-19.74$&$-16.87$&$69.24\pm 3.32$&$0.3330\pm 0.0586$\\
		\Xcline{1-8}{0.2pt}
		&Data1&$\infty$&$22.00$&$-38.89$&$-35.03$&$70.88\pm 1.65$&$0.2603\pm 0.0240$\\
		\Xcline{2-8}{0.2pt}
		$\Lambda$CDM&Data2&$\infty$&$27.45$&$-27.58$&$-23.72$&$70.65\pm 1.22$&$0.2589\pm 0.0181$\\
		\Xcline{2-8}{0.2pt}
		&Data3&$\infty$&$14.52$&$-19.50$&$-16.04$&$67.76\pm 3.09$&$0.3271\pm 0.0609$\\
		\cline{1-8}
	\end{tabular}
	\label{tab:chi2}
\end{table}

\begin{table}
	\centering
	\caption{Bayes factor}
	\begin{tabular}{ |c | c |c | }
		\hline
		Data & Bayes factor ($B_{ij}$) & $|\ln B_{ij}|$ \\  \hline
		OHD (DA) + OHD (Homogenous) & $B_{01} = 6.17$ & 1.82 \\ \Xcline{2-3}{0.2pt}
		& $B_{02} =1.75 $ & 0.55 \\ \hline
		OHD (DA) + OHD (Non-homogenous) & $B_{01} = 32.63$ & 3.48 \\ \Xcline{2-3}{0.2pt}
		& $B_{02} = 3.12$ & 1.13 \\ \hline
		OHD (DA) & $B_{01} = 0.99$ & 0.01 \\ \Xcline{2-3}{0.2pt}
		& $B_{02} = 1.02$ & 0.01 \\ \hline
		
	\end{tabular}
	\label{tab:3}
\end{table}

\section*{Cosmological parameters}
\begin{figure}
 	\centering
 	\includegraphics[width=0.65\textwidth]{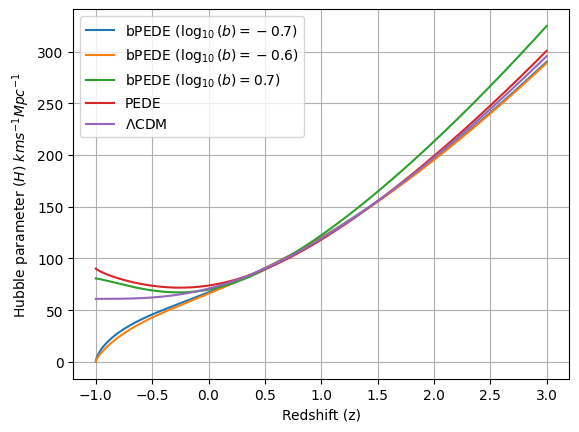}
 	\caption{Evolution of Hubble parameter ($H$) against redshift ($z$) is plotted for the best-fit bPEDE, PEDE and $\Lambda$CDM models. }
 	\label{fig:Hubble}
 \end{figure}

From the analysis presented in the last section, we have seen that the bPEDE model better fits the observational Hubble data than the standard $\Lambda$CDM and PEDE models. In this section, we present a comparative study of evolution of the various cosmographic parameters for the bPEDE, PEDE and the $\Lambda$CDM models. The rate of expansion of the universe is encoded in the Hubble parameter, which is given by Eq. (\ref{eq:H}). The evolution of the Hubble parameter against redshift for the best-fit bPEDE, PEDE and $\Lambda$CDM models are shown in Fig. \ref{fig:Hubble}. The Hubble parameter of the best-fit bPEDE with $\log_{10}(b) = -0.7, -0.6$ shows a similar behaviour where the Hubble parameter is an ever-decreasing function of the scale factor.
 Interestingly, the Hubble parameter tends to zero in the asymptotic future when $z\rightarrow -1$ or equivalently $a\rightarrow \infty$, indicating the possibility of a static universe in the asymptotic future. 
 In the asymptotic future, as $z$ approaches $-1$, the Hubble parameter stabilizes to a constant value within the $\Lambda$CDM model, manifesting a deSitter-type evolution of the universe. In the context of the PEDE and bPEDE model with $b = \log_{10}(0.7)$, the Hubble parameter shows a decreasing trend in the past, followed by an increase in the future, finally converging to a constant value as $z\rightarrow -1$. Notably, the asymptotic value in these models marginally exceeds the Hubble parameter value of the $\Lambda$CDM model. 
\begin{figure}
	\centering
	\includegraphics[width=0.65\textwidth]{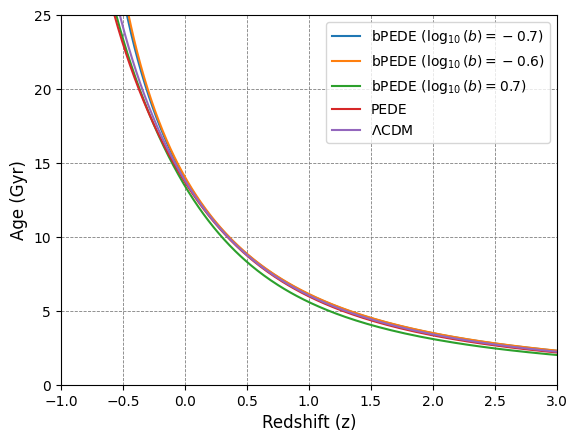}
	\caption{Evolution of age of the universe against redshift (z) is plotted for the best-fit bPEDE, PEDE and $\Lambda$CDM models. The present age of the universe is the age at $z=0$.}
	\label{fig:age}
\end{figure}
\begin{table}
	\centering
	\caption{Age ($t_0$), present value of the equation of state parameter ($\omega_{D_0}$), present value of the deceleration parameter ($q_0$) and transition redshift ($z_T$) of the universe computed for best-fit bPEDE, PEDE model and $\Lambda$CDM models.}
	\begin{tabular}{ |c | c | c | c | c | }
		\hline
		Model & $t_0$ &  $\omega_{D_0}$  & $q_0$ & $z_T$  \\  \hline
		bPEDE ($\log_{10}(b)=-0.7)$ & 13.96&  -0.79  & -0.36 &   0.79  \\ \hline
		bPEDE ($\log_{10}(b)=-0.6)$ & 14.05&  -0.76  & -0.32 &  0.78  \\ \hline
		bPEDE ($\log_{10}(b)=0.7)$ & 13.44& -1.2   & -0.70 & 0.59  \\ \hline
		PEDE & 13.71& -1.1   & -0.78 &  0.78   \\ \hline
		$\Lambda$CDM & 13.82 & -1 & -0.60 & 0.78  \\ \hline
	\end{tabular}
	\label{tab:cp}
\end{table}
The expression to compute the present age of the universe that follows from the definition of the Hubble parameter is
\begin{align}
	\label{eq:age}
	t_0 - t_B = \int_{0}^{1}\frac{1}{aH(a)}da,
\end{align}
where $t_0$ is the present age of the universe, and $t_B$ is the age of the universe at the Big Bang, which is assumed to be zero.
The age of the universe computed for bPEDE, PEDE and $\Lambda$CDM models are presented in Tab. \ref{tab:cp}. The universe's age computed for the bPEDE models with $\log_{10}(b) = -0.7, -0.6$ is slightly higher than the universe's age computed with the $\Lambda$CDM model. In comparison, the bPEDE model with $\log_{10}(b) = 0.7$ and PEDE model predict a slightly lower age than the standard model prediction where age predicted by PEDE model is closer to the $\Lambda$CDM model. 

\begin{figure}
	\centering
	\includegraphics[width=0.65\textwidth]{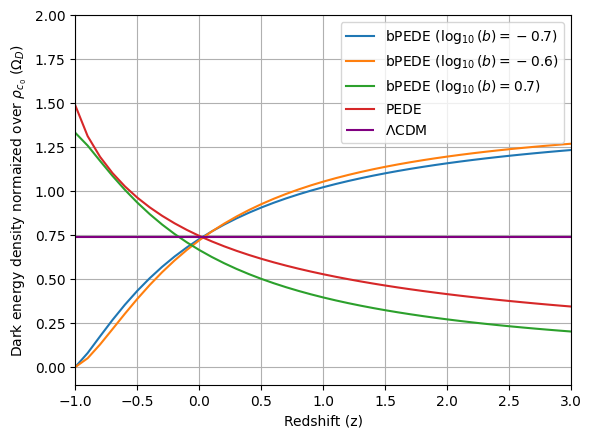}
	\caption{Evolution of dark energy density normalized over the present value of the critical density against redshift (z) is plotted for the best-fit bPEDE, PEDE and $\Lambda$CDM models.}
	\label{fig:density}
\end{figure}

The evolution of the matter density has the same form for all the models, i.e. $\Omega_m = \Omega_{m_0}(1+z)^3$. The matter density decreases with the increase in scale factor and approaches zero in the asymptotic future $a\rightarrow \infty.$ The evolution of dark energy density normalized over the present value of the critical density ($\rho_{c_0}$) for all the models under consideration are shown in Fig. \ref{fig:density}. The dark energy density has had no effective presence in the past for the PEDE and bPEDE ($\log_{10}(b) = 0.7$) models, and the densities increase with an increase in scale factor and asymptotically reaches a value close to $2\Omega_{D_0}$. On the other hand, the dark energy densities within the  bPEDE models ($\log_{10}(b) = -0.7, -0.6$) show distinct behaviour where the dark energy density has the value $2\Omega_{D_0}$ in the past and asymptotically tends to zero in the far future. However, the dark energy density dominates over the matter density at present for all the models, indicating that all the models successfully explain the late-phase accelerated expansion of the universe.

\begin{figure}
	\centering
	\includegraphics[width=0.65\textwidth]{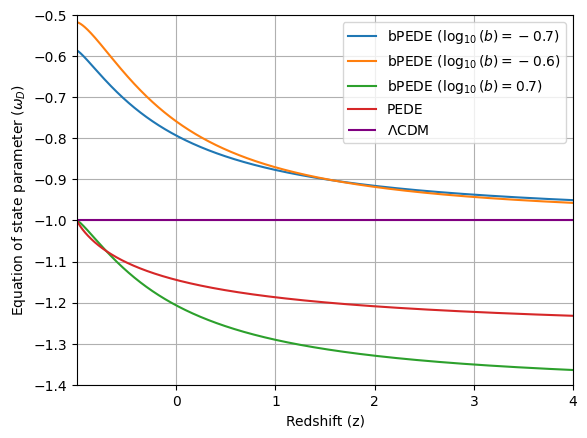}
	\caption{Evolution of equation of state parameter of dark energy against redshift (z) is plotted for the best-fit bPEDE models, PEDE model and $\Lambda$CDM model. The present $\omega_D$ corresponds to the value at $z=0$.}
	\label{fig:omega}
\end{figure}

The progress of the equation of state parameter of the dark energy density with the redshift for best-fit bPEDE models, PEDE model and the $\Lambda$CDM model, are shown in Fig. \ref{fig:omega}.
From Fig. \ref{fig:omega}, it is evident that the value of $\omega_D$ is in between -1/3 and -1 throughout the evolution of the universe for the bPEDE ($\log_{10}(b) = -0.7, -0.6$) models, resembles the quintessence nature of dark energy whereas the $\omega_D > -1$ throughout the evolution of the universe for the bPEDE ($\log_{10}(b) = 0.7$) and PEDE models, resembles the phantom nature of dark energy. The $\omega_D$ tend to -1 in the far past for the quintessence-type bPEDE models, while the $\omega_D$ tend to -1 in the asymptotic future for the phantom-type bPEDE and PEDE models. The present values of the equation of state of dark energy, $\omega_{D_0}$ for all the models are given in Tab. \ref{tab:cp}. The present $\omega_D$ is slightly less than -1 for quintessence-type bPEDE models, while the values are slightly higher than -1 for the Phantom-type bPEDE and PEDE models.

\begin{figure}
	\centering
	\includegraphics[width=0.65\textwidth]{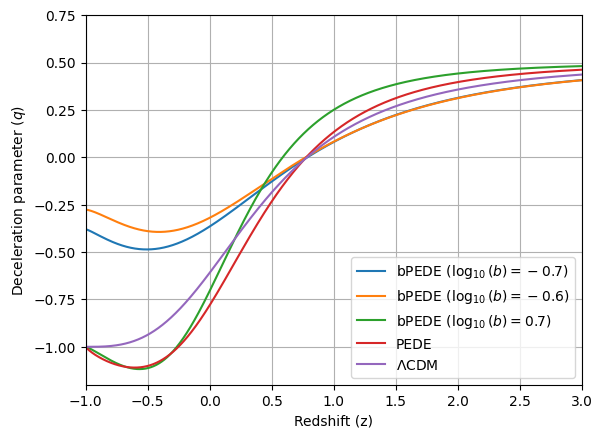}
	\caption{Evolution of deceleration parameter of the universe against redshift (z) is plotted for the best-fit bPEDE, PEDE and $\Lambda$CDM models.}
	\label{fig:deceleration}
\end{figure}

Evolution of the deceleration parameter ($q$) of the universe with respect to redshift ($z$) for all the models are shown in Fig. \ref{fig:deceleration}. The present values of the deceleration parameter and the transition redshift ($z_T$) for all the models are summarized in Tab. \ref{tab:cp}.
From the Fig. \ref{fig:deceleration}, it is clear that all the models under consideration predict the decelerating to accelerating transition, and the present value of the deceleration parameter is negative for all the models showing that the present universe is undergoing an accelerating expansion. The decelerating to accelerating phase transition occurred at a redshift $z_T \sim 0.78$ for quintessence-type bPEDE, PEDE and the $\Lambda$CDM models while the $z_T$ is slightly less value for the phantom-type bPEDE model. 
\begin{figure}
	\centering
	\includegraphics[width=0.65\textwidth]{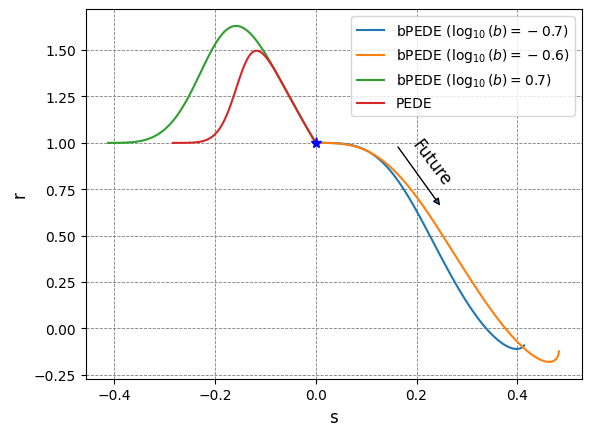}
	\caption{The r-s trajectory is plotted for the best-fit bPEDE, PEDE and $\Lambda$CDM models. The $r=1$, $s = 0$ (blue star) is a fixed point for the standard $\Lambda$CDM model.}
	\label{fig:rs}
\end{figure}

The Statefinder, originally proposed by Sahni et al. \cite{sahni2003statefinder}, is a geometric diagnostic tool that distinguishes between dark energy models. In statefinder analysis, the geometric pair jerk parameter (r) and the snap parameter (s) characterize the dark energy models. The jerk parameter is defined as
\begin{align}
	\label{eq:state}
	r = \frac{1}{aH^3}\frac{d^3a}{dt^3} \hspace{1cm}
	s = \frac{r-1}{3(q-1/2)}
\end{align}
where $a$ is the scale factor, $H$ is the Hubble parameter and $q$ is the deceleration parameter. It is convenient to express $r$ and $s$ in terms of derivative with respect to the parameter $x$, where $x = \ln a$. Then $r$ and $s$ can be expressed as
\begin{align}
	\label{eq:r}
	r = \frac{1}{2h^2}\frac{d^2h^2}{dx^2} + \frac{3}{2h^2}\frac{dh^2}{dx} + 1 ,
\end{align}
\begin{align}
	\label{eq:s}
	s = -\left(\frac{\frac{1}{2h^2}\frac{d^2h^2}{dx^2} + \frac{3}{2h^2}\frac{dh^2}{dx}}{\frac{3}{2h^2}\frac{dh^2}{dx} + \frac{9}{2}}\right).
\end{align}
Substituting the expression for $h = H/H_0$ from Eq. (\ref{eq:H}) in \ref{eq:r} and \ref{eq:s}, we obtain the evolution of $r$ and $s$ parameter for the bPEDE models as,
\begin{align}
	\label{eq:r1}
	r =  \frac{\Omega_{D_0}\sech^2(\log_{b}(1+z))}{(\ln b)^2h^2}\left[\tanh(\log_{b}(1+z)) + \frac{3}{2}\ln b\right] + 1,
\end{align}
\begin{align}
	\label{eq:s1}
	s = -\left[\frac{\Omega_{D_0}\sech^2(\log_{b}(1+z))\left[\tanh(\log_{b}(1+z)) + \frac{3}{2}\ln b\right]}{\frac{3}{2}(\ln b)^2\left(-3\Omega_{m_0}(1+z)^3 + \frac{\Omega_{D_0}}{\ln b}\sech^2(\log_{b}(1+z)) + 3h^2\right)}\right].
\end{align}

 We also obtain the $r$ and $s$ evolution of PEDE and $\Lambda$CDM model using the respective Hubble parameter evolution. The r-s trajectory of all the studied models is presented in Fig. (\ref{fig:rs}).
The $(r,s) = (1, 0)$ is a fixed point for the $\Lambda$CDM model. The $r>1$ and $s<0$ for the PEDE and bPEDE ($\log_{10}(b) = 0.7$) model depicting the phantom nature of dark energy density. The $r$ and $s$ values of these models reach the $\Lambda$CDM fixed point in the far future. The $r<1$ and $s>0$ for the bPEDE ($\log_{10}(b) = -0.7, -0.6$), depicting quintessence nature of dark energy. The $r$ and $s$ values of these models possess the $\Lambda$CDM fixed point in the far past. In conclusion, the models that better fit the observational data are quintessence type. Interestingly, the best-fit bPEDE models predict the present value of the Hubble parameter closely aligns with the CMBR data while inconsistent with the local measurements.
\section*{Conclusion}

Motivated by the investigation of the Hubble tension problem within the framework of Emergent Dark Energy (EDE) models as outlined in ref. \cite{li2020evidence, yang2021generalized, hernandez2020generalized, padilla2021cosmological}, we explored the possibility of Phenomenological Emergent Dark Energy (PEDE)-type models in light of observational Hubble data. Our particular interest lied in identifying the PEDE-type models that yield a better fit to the observational Hubble data as compared to the PEDE and the $\Lambda$CDM models and also investigating whether $H_0$ value aligns with the CMBR measurement or the value obtained by the local measurement. We designated this model as the bPEDE model. In the bPEDE model, the matter and dark energy are considered separately conserved. Consequently, the matter density adheres to the same evolutionary behaviour observed in the $\Lambda$CDM model. However, dark energy density is assumed to follow a specific form reminiscent of the PEDE model, $\Omega_D \propto \tanh(\log_b(1+z))$. 

In the present analysis, we considered three distinct datasets: OHD (DA), OHD (DA) + OHD (Homogenous) and OHD (DA) + OHD (Non-homogenous). Employing a comprehensive analysis based on the Akaike Information Criterion (AIC), Bayesian Information Criterion (BIC) and the Bayesian evidence shows that the bPEDE model, parameterized with $\log_{10}(b)=-0.7, -0.6$ are preferred over both the PEDE model and $\Lambda$CDM, specifically for the  OHD (DA) + OHD (Homogenous),  OHD (DA) + OHD (Non-omogenous) datasets combinations, respectively. Intriguingly, the OHD (DA) data displayed no clear preference for either model. Notably, the best-fit bPEDE models are preferred over the PEDE model with a $\Delta$AIC or $\Delta$BIC approximately $-8.1$ for the OHD (DA) + OHD (Homogenous) dataset and even more prominently at $-12.4$ for the  OHD (DA) + OHD (Non-homogenous) dataset. The same trend was also encoded in the Bayesian evidence. Interestingly, the best-fit bPEDE models predict the value of $H_0$ closely aligns with the value obtained from the CMBR measurement. Indeed, it is essential to compare the predictions of the bPEDE model with the Cosmic Microwave Background Radiation (CMBR) data to determine whether the model can solve the Hubble tension problem. Furthermore, a comprehensive analysis incorporating diverse observational probes such as Type Ia supernovae, Baryon Acoustic Oscillation (BAO) and Large-scale structure (LSS) in the distribution of galaxies are essential to ascertain if these datasets collectively converge towards a value of $H_0$ in agreement with the CMBR predicted value. Then, it would strongly suggest reconsidering the systematic uncertainties associated with the local measurement, which will be the prime focus of future work.

Further, we analyzed the evolution of cosmological parameters. The best-fit bPEDE models 
($\log_{10}(b) = -0.7, -0.6$) show decreasing Hubble parameters, approaching zero as $z\rightarrow -1$ or equivalently $a\rightarrow \infty$, hinting at a possible static future universe. The value of $\omega_D$ remains between $-1/3$ and $-1$ for both bPEDE models ($\log_{10}(b) = -0.7, -0.6$), akin to quintessence dark energy. Conversely, for bPEDE ($\log(b) = 0.7$) and PEDE models, $\omega_D$ consistently exceeds $-1$, resembling phantom dark energy. For the phantom dark energy models considered here, dark energy has had a negligible presence in the past. As the scale factor increases, density approaches a value near $2\Omega_{D_0}$. Conversely, quintessence-type bPEDE models display a distinct pattern, with the density starting at $2\Omega_{D_0}$ in the past and asymptotically approaching zero in the far future. The age of the universe computed for the quintessence-type bPEDE models is slightly greater than that computed using the $\Lambda$CDM model. Conversely, the Phantom-type bPEDE model and PEDE model predict a slightly younger age than the $\Lambda$CDM model, with age predicted by the PEDE model being closer to the $\Lambda$CDM prediction. All models analyzed here predict a transition from decelerated to accelerated expansion, with the present value of the deceleration parameter being negative, confirming the universe's ongoing accelerating expansion. This transition occurred at a redshift $z_T \sim 0.78$ for the quintessence-type bPEDE models, the PEDE and the $\Lambda$CDM models. However, for the phantom-type bPEDE model, the transition occurred at a slightly lower value. The statefinder analysis revealed that $r>1$ and $s<0$ for the PEDE and bPEDE ($\log(b) = 0.7$) model, depicting the phantom nature of dark energy density whereas $r<1$ and $s>0$ for the bPEDE models ($\log_{10}(b) = -0.7, -0.6$), depicting quintessence nature of dark energy.

In summary, our analysis based on observational Hubble data indicates a substantial preference for the quintessence-type bPEDE model over both the PEDE model and the standard $\Lambda$CDM model. Remarkably, the best-fit quintessence-type bPEDE model predicts a value of $H_0$ that closely aligns with the value obtained from the CMBR measurements assuming the $\Lambda$CDM model. Furthermore, in contrast to the $\Lambda$CDM model and PEDE model, the quintessence-type bPEDE models predict the static future for the universe.

%\bibliography{ref.bib}
%\bibliographystyle{ieeetr}

\end{document}